\documentclass[prl,twocolumn,preprintnumbers,amsmath,amssymb,floatfix]{revtex4-1}

\usepackage[dvipsnames]{xcolor}
\usepackage{soul}
\usepackage{amssymb}
\sethlcolor{green}

\usepackage{graphicx}
\usepackage{dcolumn}
\usepackage{bm,textcomp}
\usepackage{booktabs,threeparttable}
\usepackage{subfigure}
\usepackage{epstopdf}
\usepackage{color}
\makeatletter 
\def\@hangfrom@section#1#2#3{\@hangfrom{#1#2#3}}
\makeatother
\makeatletter 
\def\@biblabel#1{[#1]}
\makeatother

\newcommand{\cm}[1]{} 
\DeclareGraphicsExtensions{.eps,.mps,.pdf,.jpg,.png}
\definecolor{bg}{rgb}{0.75,.75,.67}
\graphicspath{{./}}

\usepackage{hyperref}
\begin{document}

\title{Zr-Co-Al bulk metallic glass composites containing B2 ZrCo via rapid quenching and annealing}
\author{Yu Chen$^{1}$, Chunguang Tang$^{1*}$, Kevin Laws$^{1}$, Qiang Zhu$^{2}$, Michael Ferry$^{1}$}

\affiliation{$^1$ School of Materials Science and Engineering, The University of New South Wales, NSW 2052, Australia; $^2$ Electronic Microscopy Unit, University of New South Wales, UNSW, Sydney, NSW 2052, Australia}

\begin{abstract}

As a promising remedy for overcoming the limited ductility and work softening of bulk metallic glasses (BMGs), BMG composites incorporating a B2 crystalline phase have attracted considerable attention. Here, we explore the formation of Zr-Co-Al BMG composites by quenching alloys Zr$_{55}$Co$_{31}$Al$_{14}$, Zr$_{54.5}$Co$_{33.5}$Al$_{12}$, Zr$_{53.5}$Co$_{36.5}$Al$_{10}$, Zr$_{52.5}$Co$_{37.5}$Al$_{10}$, and Zr$_{43}$Co$_{43}$Al$_{14}$. We found the first alloy fully amorphous whereas the fifth was fully crystallized upon quenching. The other three were quenched to generate composite structures, with a higher fraction of B2 ZrCo phase with increasing Co/Zr ratio and decreasing Al content. For comparison, the formation of B2 ZrCo in annealed Zr$_{55}$Co$_{31}$Al$_{14}$ was also studied. For both approaches the influence of crystalline phases on hardness was examined.

 \vspace{10pt}

Published at \href{https://doi.org/10.1016/j.jallcom.2019.153079}{Journal of Alloys and Compounds 820 (2020) 153079.}
\end{abstract}

\maketitle

\section{Introduction}

Bulk Metallic Glasses (BMGs) are promising structural materials with high strength, hardness and elastic limit. However, their low tensile ductility at failure and strain-softening associated with autocatalytic shear-banding at room temperature significantly hamper their applications. These problems have been tackled by designing BMG composites containing ductile crystalline phases dispersed throughout the glassy matrix \cite{hays_microstructure_2000}. While compressive work hardening \cite{zhang_strong_2013} and large tensile ductility \cite{hofmann_designing_2008,hays_microstructure_2000} have been reported for BMG composites with ductile crystalline phases, the combination of tensile plasticity and work hardening is only accomplished via BMG composites containing the B2 (Pm/3m, space group number 221) crystalline phase, such as Zr-Cu-Al BMG composites \cite{pauly_transformation-mediated_2010, wu_bulk_2010, kim_realization_2011, zhang_strong_2013}.  Different from conventional ductile crystalline phases, the B2 phase can respond to the applied stress or strain via martensite transformation (B2 to B19$^\prime$ or B2 to B33), resulting in work-hardening and transformation-induced plasticity. 

Recently, a few studies have shown that ZrCo in the B2 phase, due to its large number of slip systems, is inherently ductile \cite{yamaguchi_room-temperature_2005, kaneno_tensile_2008, li_transformation-induced_2013, li_mechanical_2016}. For example, tensile elongations greater than 7\% \cite{yamaguchi_room-temperature_2005} and 20\% \cite{kaneno_tensile_2008} have been reported for B2 ZrCo after hot-rolling and recrystallization,  and compressive strain greater than 40\% \cite{li_mechanical_2016} has also been reported for as-cast B2 ZrCo. On the other hand, glass forming abilities (GFAs) of Zr-Co-Al alloys have been systematically investigated \cite{zhang_optimum_2004, zhang_formation_2004, wada_formation_2009} and some alloys with high GFA have been identified. For example, fully amorphous Zr$_{56}$Co$_{28}$Al$_{16}$ cylinders with diameter up to 18 mm was produced \cite{wada_formation_2009}. Amorphous Zr$_{56}$Co$_{44-x}$Al$_{x}$ ($x$=12, 14, 16, 18) exhibits high compressive plasticity between $\sim$5\% and $\sim$11\% and strength more than 2 GPa \cite{tan_study_2011, tan_formation_2012}. The success of CuZr-based BMG composites implies the potential to further improve the mechanical properties of Zr-Co-Al BMGs by designing some composites containing B2 ZrCo. There have been some attempts to produce B2 ZrCo phase within Zr-Co-Al BMG matrix by annealing \cite{tan_study_2011}, but so far as-cast Zr-Co-Al BMG composites have not been reported.  We noted that B2 ZrCo phase has been found in as-cast (ZrCo)$_{100-x}$Al$_{x}$ (x=1, 2, 3, 5), but these alloys have poor GFA and are fully crystallized upon quenching \cite{li_transformation-induced_2013, li_enhanced_2013, li_microstructure_2017}. 

In this work, by exploring the ternary composition space we identify certain compositions of significant potential for the formation of BMG composites containing B2 ZrCo. For comparison, we also study the formation of B2 ZrCo in devitrified Zr$_{55}$Co$_{31}$Al$_{14}$. 

\section{Methods}

\begin{figure}[ht]
\includegraphics[width=3.3in]{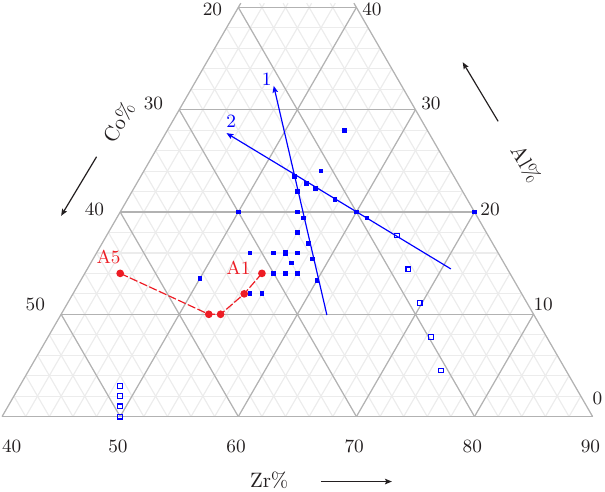}
\caption{Zr-Co-Al composition diagram. The  square symbols (solid for amorphous and open for crystal) represent some previously explored compositions of various sample size \cite{wada_formation_2009, wada_formation_2002, tan_formation_2012, zhang_optimum_2004,zhang_formation_2004,li_transformation-induced_2013, li_enhanced_2013, li_microstructure_2017,hua_ni-_2011}. Arrows 1 and 2 indicate two series of compositions studied in reference \cite{zhang_optimum_2004}, which have increasing and  constant conduction electron concentration, respectively. The red circles represent the compositions in this work, and the crystal phases in the samples are listed in a descending order of their X-ray diffraction intensities. The dashed line is for eye guide.}
\label{fig:fig1}
\end{figure}

To date a number of Zr-Co-Al compositions have already been explored, as indicated by the solid blue squares (for high GFA) or the open blue squares (for poor GFA) in Fig. \ref{fig:fig1}. We start with a composition near the glass-forming Zr$_{56}$Co$_{28}$Al$_{16}$ and increase its crystallization ability by changing the composition towards Zr$_{50}$Co$_{50}$. To this end we selected three compositions, Zr$_{55}$Co$_{31}$Al$_{14}$, Zr$_{54.5}$Co$_{33.5}$Al$_{12}$, and Zr$_{53.5}$Co$_{36.5}$Al$_{10}$. A fourth alloy, Zr$_{52.5}$Co$_{37.5}$Al$_{10}$, was also selected in view of the relatively poor GFA for compositions with Al below 10\%. Since alloy (ZrCo)$_{100-x}$Al$_{x}$ is well characterized by B2 ZrCo but has poor GFA at low $x$, we also studied alloy Zr$_{43}$Co$_{43}$Al$_{14}$. For simplicity, these five alloys are denoted thereafter as alloys A1 to A5, respectively. All five compositions studied in this work are indicated in Fig. \ref{fig:fig1} by the solid red circles.

\begin{figure}[ht]
\includegraphics[width=3.2in]{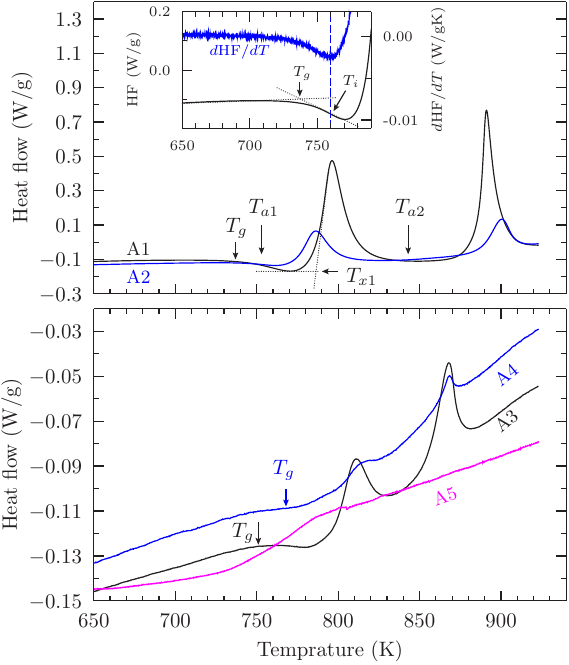}
\caption{DSC curves for as-cast alloys A1-A5, and the obtained thermal properties shown in Table \ref{tab:temp}. $T_{a1}$ and $T_{a2}$ indicate the annealing temperatures for alloy A1. Inset in top panel: the determination of $T_g$ of alloy A1. $T_i$ is the inflection point where the heat flow has the local lowest derivative ($d$HF/$dT$).}
\label{fig:dsc}
\end{figure}

\begin{figure}[ht]
\includegraphics[width=3.3in]{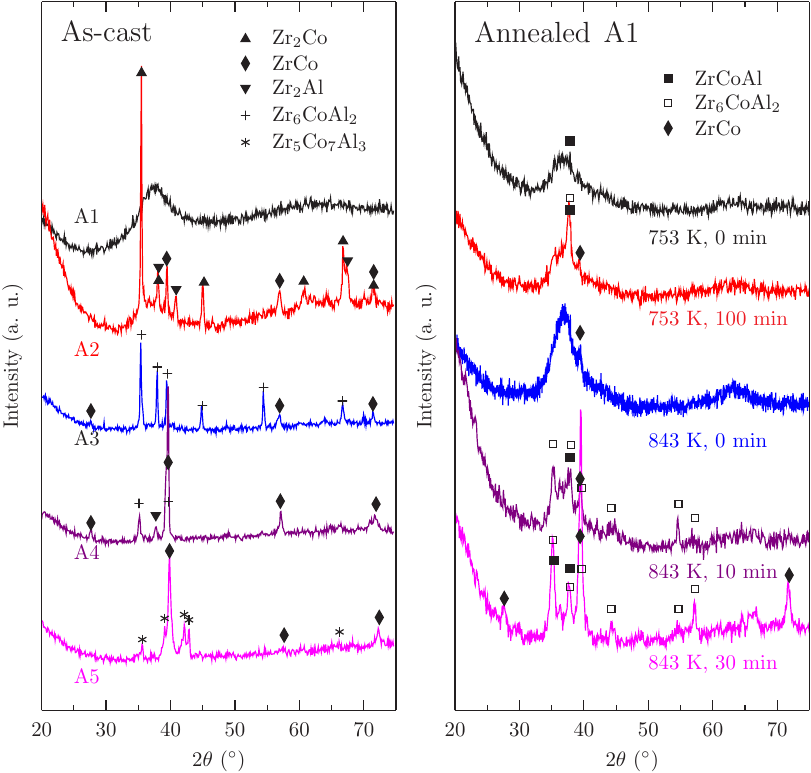}
\caption{(left) XRD patterns for as-cast alloys A1-A5; (right) alloy A1 annealed at 753 K and 843 K for various time. }
\label{fig:xrd}
\end{figure}

Alloys were fabricated by mini arc melting the mixture of pure Zr (99.95\%), Co (99.95\%) and Al (99.99\%) metals in a Ti-gettered high-purity argon atmosphere. The alloys were remelted a minimum of six times before being suction-cast into cylinders of 3 mm in diameter and 40 mm in length. Thermal analysis was performed using differential scanning calorimetry (DSC, NETZSCH 404) at a heating rate of 20 K/min. Structural characterizations were performed using X-ray diffraction (XRD, Bruker D8, Philips X’Pert, Operating Voltage of 45 kV and Cu-K$_\alpha$ radiation), Scanning Electron Microscopy (SEM, Hitachi 3400X) attached with an Energy dispersive Spectroscopy (EDS). Micro-Hardness was measured using a Struers Durascan micro-hardness tester at 1 kg with dwell time of 15 seconds. To examine the effect of devitrification on the formation of the B2 phase, Zr$_{55}$Co$_{31}$Al$_{14}$ was annealed using DSC near the onsets of the first and second crystallization, respectively, for various times before cooling the samples by switching off the DSC. The heating rate for annealing was 20 K/min. 

\section{Results and Discussion}
\begin{table*}
\tabcolsep=11pt
\caption{Thermal and electronic properties of alloys A1-A5 and similar reported alloys. The glass transition temperature ($T_g$), the onset temperature of crystallization ($T_x$), the width of the supercooled liquid region ($\Delta T_x$), and the latent heat of crystallization ($\Delta H$) were measured at a heating rate of 20 K/min, same as those in reference \cite{tan_formation_2012}. The average number of effective conduction electrons per atom ($e_c/a$) is also listed.}
\begin{tabular}{l l l l l l l l}
\hline \hline
alloy& $T_g$ (K)& $T_{x1}$ (K) & $\Delta T$ (K) & $\Delta H_1$ (J/g) & $\Delta H_2$ (J/g) & e$_c$/a & reference\\  
\hline
Zr$_{55}$Co$_{31}$Al$_{14}$ (A1)&  737& 785 & 49 & 33.64 & 27.59 &1.245 & this work\\
Zr$_{54.5}$Co$_{33.5}$Al$_{12}$ (A2) &737  & 774 & 37 & 10.38 & 7.58 &1.178 &this work \\
Zr$_{53.5}$Co$_{36.5}$Al$_{10}$ (A3) & 751 & 798 & 37 & 1.28 & 1.53 &1.103 &this work \\
Zr$_{52.5}$Co$_{37.5}$Al$_{10}$ (A4) & 768 & 803 & 35 & 0.15 & 0.27 &1.088 &this work \\
Zr$_{43}$Co$_{43}$Al$_{14}$ (A5) & \multicolumn{5}{c}{in as-cast crystal state} & 1.065&this work \\
Zr$_{56}$Co$_{30}$Al$_{14}$ &  748& 788 & 40& - & - & 1.26& \cite{tan_formation_2012}\\
Zr$_{56}$Co$_{32}$Al$_{12}$ &  747& 785 & 38& - & - & 1.20& \cite{tan_formation_2012}\\
\hline\hline
\end{tabular}
\label{tab:temp}
\end{table*}

The DSC curves of as-cast alloys A1-A5 are shown in Fig. \ref{fig:dsc}, with the thermal parameters summarized in Table \ref{tab:temp}. As can be seen, as the content of Al decreases and Zr/Co ratio increases in alloys A1-A4, the latent heat associated with the two crystallization peaks decreases. For alloy A5, no crystallization peaks were detected. These results indicate decreasing GFA as the composition changes from alloy A1 to A5. The width of supercooled liquid region ($\Delta T_x{=}T_{x1}{-}T_g$) for alloys A1 and A2 was determined to be 49 and 37 K, respectively. These are close to the reported values for similar compositions Zr$_{56}$Co$_{30}$Al$_{14}$ (40 K) and Zr$_{56}$Co$_{32}$Al$_{12}$ (38 K).

Based on the proposal of Nagel and Tauc \cite{nagel_nearly-free-electron_1975}, whereby metallic glasses can be stabilized at compositions for which the Fermi level is located at a minimum electronic density of states, some authors \cite{zhang_optimum_2004} have investigated the GFA of Zr-Co-Al alloys from the perspective of conduction electron concentration, or the number of (effective) conduction electrons per atom ($e_c/a$). Using the assigned elemental $e_c/a$ values (1.5, 0, and 3 for Zr, Co, and Al, respectively \cite{wang_composition_2003}), they found a positive correlation between the stability of glass and $e_c/a$ in the range of 1.3${<}e_c/a{<}$1.5 for (Zr$_9$Co$_4$)$_{100-x}$Al$_x$, as indicated by arrow 1 in Fig. \ref{fig:fig1}. In our study, $e_c/a$ decreases from 1.245 for alloy A1 to 1.088 for alloy A5 (Table \ref{tab:temp}), which appears to be consistent with the decreasing GFA and the previous findings \cite{zhang_optimum_2004}. However, it is worth noting that Zr$_{64.6}$Co$_{17.7}$Al$_{17.7}$ ($e_c/a$=1.5) and some other compositions have high $e_c/a$ values but poor GFA \cite{zhang_optimum_2004} indicating that there may be a fixed e/a band for glass formation, or other effects are at play in these alloys.

It is interesting to compare the GFA of (ZrCo)$_{100-x}$Al$_x$ and its counterpart (ZrCu)$_{100-x}$Al$_x$ for low $x$ since both Zr$_{50}$Co$_{50}$ and Zr$_{50}$Cu$_{50}$ can exhibit B2 crystal phase, which, according to the binary phase diagrams, precipitates directly from the liquid. Previous results \cite{li_transformation-induced_2013, li_enhanced_2013, li_microstructure_2017} and the present work indicate that (ZrCo)$_{100-x}$Al$_x$ has poor GFA for $x$ as high as 14, but (ZrCu)$_{100-x}$Al$_x$ ($x\leq10$) is a good glass former \cite{kumar_plasticity_2007}, albeit with nanocrystals found in the glass matrix \cite{das_work-hardenable_2005, kim_heterogeneity_2006, zhang_-situ_2008}. This is consistent with the fact that Cu contributes one (effective) electron \cite{chen_bulk_2003} and hence less Al is required for a large glass.  The significant GFA difference could also be a reflection of the stability of B2 ZrCo and B2 ZrCu, the former being stable at low temperature while the latter decomposing into Cu$_{10}$Zr$_7$ and CuZr$_2$ around 1000 K according to the phase diagram \cite{okamoto_cu-zr_2012}, although we note that the ternary phase equilibria of the systems is a little different, with many ternary intermetallics in the Zr-Co-Al system being present.  

As GFA decreases, the alloys (partially) crystallize during quenching. As shown in Fig. \ref{fig:xrd}, three crystalline phases, namely Zr$_2$Co, ZrCo and Zr$_2$Al, were detected in alloy A2. The formation of Zr$_2$Co and ZrCo phases is understandable since, if we neglect the influence of Al, binary alloy Zr$_{62}$Co$_{38}$, corresponding to the Zr/Co ratio of alloy A2, solidifies into Zr$_2$Co and ZrCo. For alloys A3 and A4, which have lower Zr/Co ratio, crystal phases Zr$_2$Co and Zr$_2$Al disappear and a new Zr$_6$CoAl$_2$ phase forms. At the same time, the fraction of B2 ZrCo phase increases as the composition changes from alloy A2 to A4. Alloy A5 was fully crystallized                                                                                                                                                                                                                                          on quenching, characterized by a dominant B2 ZrCo phase and a small amount of Zr$_5$Co$_7$Al$_3$ phase. These two phases are also those found in as-cast Zr$_{47.5}$Co$_{47.5}$Al$_5$ \cite{li_transformation-induced_2013}, where the Zr$_5$Co$_7$Al$_3$ phase segregates to the grain boundaries of the B2 ZrCo phase. For 3\% or lower Al content, the influence of Al on crystallization dynamics of (ZrCo)$_{100-x}$Al$_x$ is negligible and the crystalline phases are B2 ZrCo, Co$_2$Zr, and CoZr$_2$ \cite{li_microstructure_2017}. 

\begin{figure*}[ht]
\includegraphics[width=5.5in]{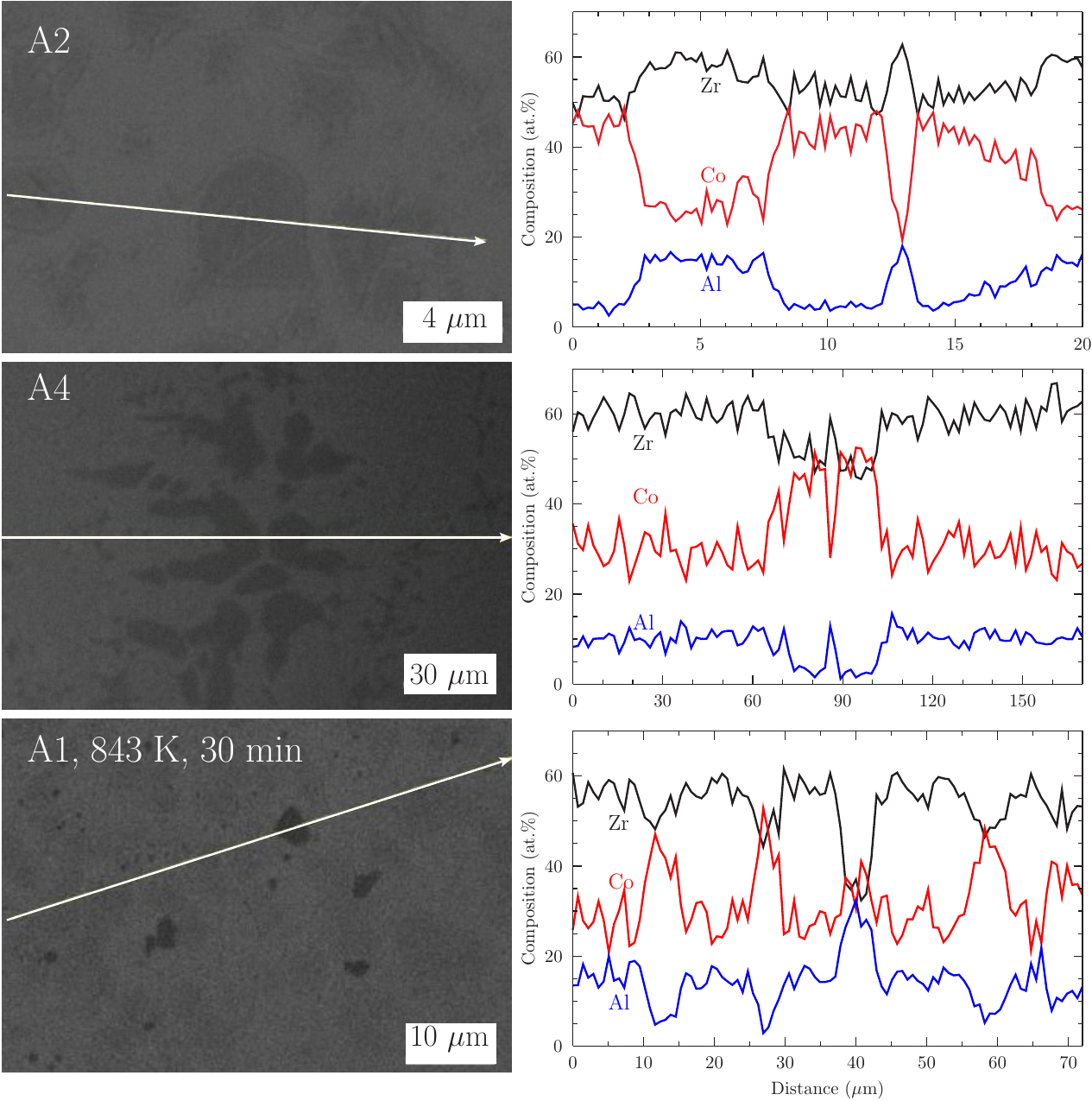}
\caption{SEM pictures (left) and the corresponding line-scanning compositions (right) for selected alloys. For as-cast alloys A2 and A4, crystalline areas with composition near Zr$_{50}$Co$_{50}$ were found. For alloy A1 annealed for 30 min at 843 K, crystalline areas with composition close to Zr$_{50}$Co$_{50}$ and ZrCoAl were found.}
\label{fig:sem}
\end{figure*}

\begin{figure}[ht]
\includegraphics[width=3.3in]{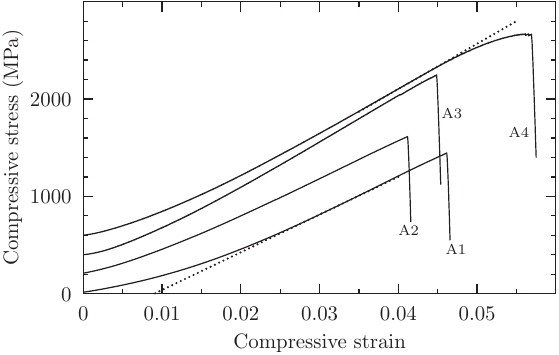}
\caption{Uncorrected compressive stress strain curves for as-cast samples. The curves are shifted vertically for visional clarity. The linear dotted lines for A1 and A4 are for eye guide.}
\label{fig:compression}
\end{figure}

\begin{figure}[ht]
\includegraphics[width=3.3in]{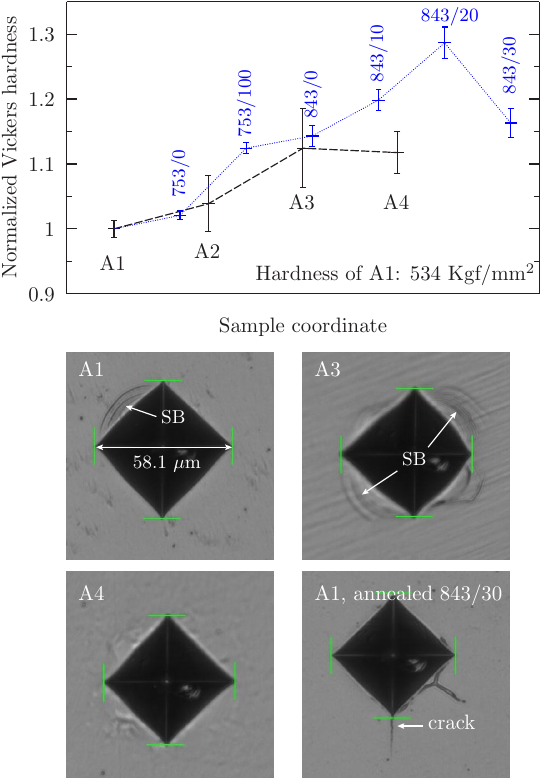}
\caption{(Top): Vickers hardness of as-cast alloys and annealed alloy A1, normalized by that of alloy A1. Label `753/0' means annealing at 753 K with zero dwell time. (Bottom): Optical microscopy images of typical indents.}
\label{fig:hardness}
\end{figure}

The DSC and XRD results discussed earlier clearly indicate that $in$-$situ$ Zr-Co-Al BMG composites can be fabricated by tuning the compositions. For comparison we also explored the formation of BMG composites by annealing alloy A1 for various dwell times at 753 and 843 K, respectively. These two temperatures are slightly below the crystallization onsets (Fig. \ref{fig:dsc}). Transient annealing at 753 K results in an emerging peak of ZrCoAl phase, and, after annealing for 100 min, ZrCo and Zr6CoAl2 peaks also emerge, but overall the amorphous hump is still preserved. Annealing at 843 K makes the crystalline peaks more pronounced, but no new crystalline phases were observed. These three phases were observed for annealing Zr$_{56}$Co$_{28}$Al$_{16}$ and Zr$_{55}$Co$_{20}$Al$_{25}$ \cite{tan_study_2011, wada_formation_2002} and also were the phases of as-cast crystallized Zr$_{64.6}$Co$_{17.7}$Al$_{17.7}$ \cite{zhang_optimum_2004}. However, comparing the annealed and as-cast alloys in this study, the product phases are somewhat different, reflecting the complexity of crystallization in Zr-Co-Al alloys.

Fig. \ref{fig:sem} shows some typical SEM micrographs of the crystallites, together with the line scanning results. In alloys A2 and A4, dendrites with a composition close to ZrCo formed due to the segregation of Co from the matrix into the dendrites and that of Zr and Al in the opposite way. It can be seen that the dendrites in alloy A4 are larger than those in alloy A2, consistent with the DSC results. For alloy A1 annealed for 30 minutes at 843 K, areas of compositions ZrCo and ZrCoAl can be seen in the line scanning plot, which are consistent with the XRD results.

To understand the effect of BMG composite formation on mechanical properties we carried out uniaxial compression testing on the as-cast alloys A1 to A4. It can be seen from Fig. \ref{fig:compression} that the stress-strain curves are slightly curved, indicating varying Young's modulii $E$, in the initial elastic regime. We attribute this to slightly non-parallel faces of the cylindrical samples, which can result in a continuously changing $E$ at the initial deformation stage that is lower than the true $E$ (see the Appendix). In our study we also expect the influence on $E$ from machine compliance and other factors. In view of these and the fact that $E$ is relatively insensitive to small composition changes, we did not `rectify' the stress-strain curves by scaling. Comparing the measured $E$ of A1, $\sim$40 GPa based on the linear portion in Fig. \ref{fig:compression}, with the reported $\sim$80-90 GPa for Zr$_{56}$Co$_{30}$Al$_{14}$ \cite{tan_formation_2012}, we estimated that the machine compliance causes about 50\% reduction of $E$. 

For amorphous Zr$_{56}$Co$_{44-x}$Al$_x$ ($x$ from 12 to 18) \cite{tan_formation_2012}, good compressive plasticity has been reported, albeit sensitive to the fraction of free volume and processing parameters \cite{tan_correlation_2011}. In this study we did not observe noticeable plasticity in amorphous sample A1. According to Fig. \ref{fig:compression}, samples A1 and A2 have similar strengths, and samples A3 and A4 have noticeable higher strengths than A1 and A2. Compared with the other samples, A4 exhibits noticeable plasticity and hardening, which we attribute to the B2 phase, the dominant crystal phase in A4. 

We also performed Vickers hardness tests, as a complimentary approach for understanding mechanical behaviour. As shown in Fig. \ref{fig:hardness}, the hardness of as-cast alloy A1 is about 530 kgf/mm$^2$, and it increases to about 550 and 600 kgf/mm$^2$ for alloys A2 and A3, respectively. Alloy A4 has similar hardness as alloy 3A. Since the alloys in this study were quenched in the same manner, the influence of quenching rate differences can be neglected and the hardness is largely affected by the formation of crystalline phases. Consistent with our study, numerous studies \cite{zhang_influence_2005, lashgari_stress-relaxation_2018, chen_crystallization_2002, dmowski_structural_2007} have shown that the hardness of BMGs increases as the volume fraction of crystalline phases increases. Nevertheless, the influence of crystalline phases on hardness is complex and depends on various factors, such as the strength/hardness of the crystal phases relative to that of the glassy matrix, whether or not the crystalline phases strain hardens, and how their distribution interacts with shear banding. The increased hardness of BMG composites does not necessarily mean higher hardness of the embedded crystals compared with the matrix. Indeed, it has been reported for Zr-Cu-based composites \cite{wu_bulk_2010, das_designing_2009} that B2 ZrCu is softer than the glassy matrix, but the transformation-induced B19$^\prime$
ZrCu is harder than the matrix \cite{wu_bulk_2010}.

Annealing of alloy A1 also results in an increase in hardness (Fig. \ref{fig:hardness}). For annealing at 753 K and transient annealing at 843 K, the magnitude of the hardness increase is similar to that of the as-cast alloys. The maximum hardness is reached for annealing for 20 minutes at 843 K, and further annealing to 30 min results in a significant drop in hardness. It is worth noting that, while for the as-cast alloys the hardness increases because of the crystalline phase, this is not the case for transient annealing at 753 K where the crystal fraction is minimal. The increase in hardness associated with annealing near the glass transition temperature is usually attributed to free volume annihilation \cite{castellero_critical_2007, seiffodini_effects_2016} associated with structural relaxation. Nevertheless, a series of recent simulations \cite{zhang_effects_2014, peng_structural_2011, tang_formation_2016, tang_atomistic_2018} imply that the relaxation-induced increase in hardness could be a result of increased five-fold topological symmetry in annealed BMGs. Consistent with these simulations, it was found that low temperature annealing results in, along with densification, more icosahedral clusters \cite{zhang_effects_2014}, which are high in five-fold symmetry. Microscopically, the fraction of five-fold symmetry, instead of the volume, of an atomic cluster was found to positively correlate with the resistance to shearing \cite{peng_structural_2011, tang_formation_2016, tang_atomistic_2018}. 

As can be seen from Fig. \ref{fig:hardness}, shear bands formed around the indent for alloys A1 and A3, but were not observed for alloy A4 in which the fraction of glass was small according to the DSC results. This implies that shear banding becomes more difficult in the presence of crystals within the glassy matrix. A similar trend was found for the annealed A1 sample. It has been reported that extended annealing time at high temperatures can result in embrittlement, exemplified by crack formation upon indentation \cite{zhang_influence_2005}. For A1 annealed for 30 minutes at 843 K, radial surface cracks were found around one of the indentation tips, where high tensile stress exists \cite{dias_computer_2006}. The formation of cracks and a concomitant decrease in hardness for A1 indicates that long-time annealing at a high temperature, which may cause full crystallization rather than BMG composites, is not favorable for mechanical property optimization. 

For some final comments, we compare the formation of B2 phase in Zr-Co-based and Zr-Cu-based BMG composites. Due to their high GFA of Zr$_{50}$Cu$_{50}$, Zr-Cu-based BMG composites are often made with compositions close to Zr$_{50}$Cu$_{50}$ \cite{wu_bulk_2010, pauly_transformation-mediated_2010}, and, as a result, often only the B2 ZrCu phase, in spherical shape, precipitates out of the glassy matrix. This can lead to significant hardening and large tensile plasticity \cite{wu_bulk_2010}. In contrast, the formation of Zr-Co-Al BMG composites with a large fraction of equimolar Zr and Co is difficult due to the poor GFA of Zr$_{50}$Co$_{50}$. On the other hand, for Zr-Co-Al alloys with different Zr and Co concentrations, the formation of B2 ZrCo is often accompanied by other crystal phases whose effects on mechanical properties could be complex. It is also important to note that for increasing GFA of Zr-Co-Al alloys a relatively high content of Al is required, and the Al dissolved in the B2 phase (see Fig. \ref{fig:sem}) may affect the morphology and properties of the latter. In the future, it might be useful to enhance the GFA of Zr-Co-Al alloys with high fraction of equimolar Zr and Co by, for example, adding other alloying elements. 

\section{Summary}
In this work we have explored the possibility of fabricating Zr-Co-Al bulk metallic glass composites with B2 ZrCo phase via 1) rapidly quenching alloys of various compositions and 2) annealing the as-cast amorphous alloys. In the first approach, it was found that, as composition changes from Zr$_{55}$Co$_{31}$Al$_{14}$ towards Zr$_{52.5}$Co$_{37.5}$Al$_{10}$ (with increasing Co/Zr ratio and decreasing Al fraction), glass forming ability decreases and B2 ZrCo phase forms, along with Zr$_{2}$Co, Zr$_{2}$Al, and  Zr$_{6}$CoAl$_{2}$ phases. In particular, the B2 ZrCo phase was dominant in Zr$_{52.5}$Co$_{37.5}$Al$_{10}$. Alloy Zr$_{43}$Co$_{43}$Al$_{14}$ with equimolar Zr and Co was found to fully crystallize into B2 ZrCo and minor Zr$_{5}$Co$_{7}$Al$_{3}$ upon quenching. In the second approach, annealing amorphous Zr$_{55}$Co$_{31}$Al$_{14}$ close to the onset $T_g$ temperature results in structural relaxation with minimal crystallization of B2 ZrCo, ZrCoAl, and Zr$_{6}$CoAl$_{2}$. A large volume fraction of these crystals was found upon annealing at 843 K, above the first DSC exothermic peak, and the alloy was almost fully crystallized after annealing for 30 min. For both approaches, the formation of crystalline phases increases hardness of the alloys, but full crystallization upon annealing results in embrittlement and a decrease in hardness. 

\section{Appendix}
\renewcommand{\thefigure}{A\arabic{figure}} 
\renewcommand{\theequation}{A\arabic{equation}}
\setcounter{figure}{0} 
\setcounter{equation}{0} 

\begin{figure}[ht]
\includegraphics[width=3.3in]{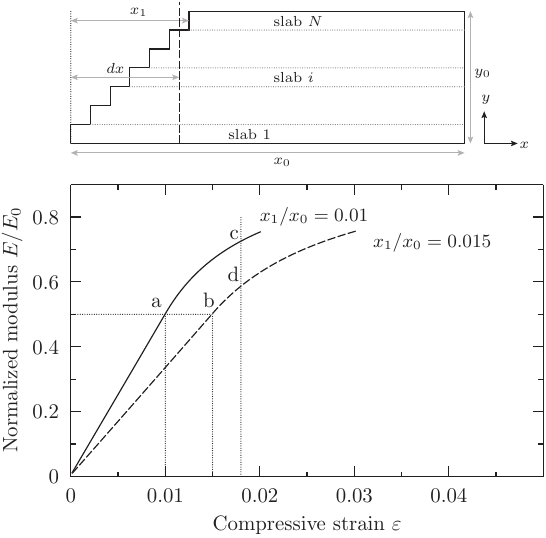}
\caption{(top) A model sample with rectangular cross-section and a non-parallel end, approximated by a number of stepped surfaces. (bottom) Effects  of strain and geometry on normalized modulus. Points $a$ and $b$ represent the strains of $x_1/x_0$, and $c$ and $d$ represent a strain beyond $x_1/x_0$.}
\label{fig:a1}
\end{figure}

To study the effect of non-parallel sample surfaces on the measured (nominal) Young's modulus, we consider a model sample as shown in Fig. \ref{fig:a1} (top). For simplicity we assume that only one surface is not normal to the loading axis $x$ and the sample cross-section is rectangular. We also assume a perfect testing machine with zero compliance. The non-parallel sample can be approximated as a combination of $N$ slabs with different lengths whose interactions are neglected. When the sample is compressed by $dx$, the strain for slab $i$ is 
\begin{equation}
 \varepsilon_i=(dx-\frac{i-1}{N-1}x_1)/(x_0-\frac{i-1}{N-1}x_1)
\end{equation}
where ${x_1/(N{-}1)}$ represents the length difference between two neighbouring slabs. The compressive force in slab $i$ is
\begin{equation}
 f_i=\varepsilon_i E_0 (y_{0} /N)z
\end{equation}
where $E_0$ is the true modulus of the sample and $y_0z$ is the cross-sectional area of the sample. The total force $F$ in the sample is the summation of $f_i$. With nominal strain $\varepsilon{=}dx/x_0$ and nominal modulus $E{=}F/(y_0z\varepsilon)$ for the sample, one can obtain the normalized modulus $E_n{=}E/E_0$ as
\begin{equation}
 E_n=\sum_{i=1}^{n} \frac{\varepsilon_i}{(N\varepsilon)} 
\label{eq:a3}
\end{equation}
where $n$ (up to $N$) is the number of slabs deformed. Several observations can be made from equation \ref{eq:a3}. First, geometrically, the measured modulus only depends on the dimensions ($x_0$ and $x_1$) along the loading axis, not the cross-sectional dimensions $y_0$ and $z$. Second, when the nominal strain $\varepsilon$ increases while the sloped end is only partially deformed ($n{<}N$) , not only $\varepsilon_i$ but also $n$ increase, resulting in a relatively rapid increase in the nominal $E$. Third, even when all the sloped surface is deformed ($n{=}N$), the nominal modulus $E$ is still smaller than the true modulus $E_0$ since $\varepsilon_i{<}\varepsilon$ for $i{>}1$. 

By assuming a large N (N=100, so that the sloped surface is `smooth'), we examined the effects of strain and geometry on the normalized modulus $E_n$. As can be seen from Fig. \ref{fig:a1}, for a certain $x_1/x_0$ value, $E_n$ increases nearly linearly as $\varepsilon$ increases up to $x_1/x_0$, and this increasing modulus causes the curvature of the stress-strain curves in the initial elastic regime (see Fig. \ref{fig:compression}). After the point $\varepsilon{=}x_1/x_0$, $E_n$ increases asymptotically towards some value less than unity (before the sample yields), and as a result the stress-strain curves become more linear at larger strains. For a given strain $\varepsilon$, the larger the $x_1/x_0$ value, the lower the nominal modulus $E$.

A previous experiment shows that a slanted contact due to a tilted platen can also cause the reduced measured modulus \cite{liu_correction_2015}. This study reveals that the measured $E$ when the sample surface is fully contacted is higher for a higher length/diameter ratio. This is consistent with our analysis since, for given diameter and tilt angle (which is equivalent to a given $x_1$ in our analysis), a high length/diameter ratio translates into a low $x_1/x_0$ and, hence, high $E$, as indicated by points $c$ and $d$ in Fig. \ref{fig:a1}.

\section{References}

\end{document}